\documentclass[aip,
amsmath,amssymb,
reprint,
floatfix,
]{revtex4-1}
\usepackage{graphicx,color}
\usepackage{dcolumn}
\usepackage{amsmath}
\usepackage{amssymb}
\usepackage{dsfont}
\usepackage[dvipsnames]{xcolor}
\usepackage{ulem}
\usepackage{siunitx}
\usepackage{bm}
\usepackage[utf8]{inputenc}
\usepackage[T1]{fontenc}
\usepackage{mathptmx}

\usepackage[hypcap=true]{subcaption}

\usepackage{hyperref}
\hypersetup{colorlinks=true,
linkcolor=blue,
breaklinks=true,
bookmarksnumbered=true,
debug=true,
naturalnames=false,
citecolor=blue,
pdfview=FitH,
pdfstartview=FitH,
}

\begin{document}

\thispagestyle{myheadings}

\title{A new stress dilatometer and measurement of the thermal expansion under uniaxial stress of Mn$_3$Sn} 

\author{Muhammad Ikhlas}
\thanks{These authors contributed equally to this work.}
\email{kent.shirer@cpfs.mpg.de}
\affiliation{Institute for Solid State Physics, University of Tokyo, Kashiwa 277-8581, Japan}
\author{Kent R. Shirer}
\thanks{These authors contributed equally to this work.}
\email{kent.shirer@cpfs.mpg.de}
\affiliation{Max Planck Institute for Chemical Physics of Solids, 01187 Dresden, Germany}
\author{Po-Ya Yang}
\affiliation{Max Planck Institute for Chemical Physics of Solids, 01187 Dresden, Germany}
\author{Andrew P. Mackenzie}
\affiliation{Max Planck Institute for Chemical Physics of Solids, 01187 Dresden, Germany}
\author{Satoru Nakatsuji}
\affiliation{Institute for Solid State Physics, University of Tokyo, Kashiwa 277-8581, Japan}
\affiliation{CREST, Japan Science and Technology Agency, Saitama 332-0012, Japan}
\affiliation{Trans-scale Quantum Science Institute, University of Tokyo, Tokyo 113-0033, Japan}
\affiliation{Department of Physics, University of Tokyo, Tokyo 113-0033, Japan}
\author{Clifford W. Hicks}
\affiliation{Max Planck Institute for Chemical Physics of Solids, 01187 Dresden, Germany}


\date{\today}

\begin{abstract}
We present a method for measuring thermal expansion under tunable uniaxial stresses, and show measurements of the thermal expansion of Mn$_3$Sn, a room temperature antiferromagnet that exhibits a spontaneous Hall effect, under uniaxial stresses of up to 1.51 GPa compression. Measurement of thermal expansion provides thermodynamic data about the nature of phase transitions, and uniaxial stress provides a powerful tuning method that does not introduce disorder. Mn$_3$Sn exhibits an anomaly in its thermal expansion near $\sim$270 K, associated with a first-order change in its magnetic structure. We show this transition temperature is suppressed by 54.6 K by 1.51 GPa compression along [0001]. We find the associated entropy change at the transition to be $\sim$\SI[mode=text]{0.1}{J.mol^{-1}.K^{-1}} and to vary only weakly with applied stress. 
\end{abstract}


\maketitle


The application of uniaxial stress modifies lattice constants and can provide powerful insight into the electronic structure of materials. Its effects often differ qualitatively from those of hydrostatic stress. For example, the superconducting critical temperature of YBa$_2$Cu$_3$O$_{6.67}$ is enhanced by hydrostatic stress, but suppressed by in-plane uniaxial stress.\cite{Kim_2018a} To date, several techniques have been combined with uniaxial stress, including magnetic susceptibility,\cite{Steppke_2017a} transport,\cite{Barber_2018a} nuclear magnetic resonance,\cite{Kissikov_2018a} and muon spin rotation.\cite{Grinenko_2020a} Dilatometry, the measurement of the thermal expansion of a material, is also an attractive technique to apply to uniaxially stressed samples, because it provides extremely high-precision bulk thermodynamic information. It is orders of magnitude more sensitive to changes in lattice parameter than x-ray or neutron diffraction.\cite{Meingast_1990a} It has, for example, been utilized in measurements of electron-lattice coupling in electronically and magnetically ordered phases\cite{Meingast_1990a, Meingast_1991a} and to probe thermodynamic effects of quantum criticality.\cite{Westerkamp_2009a, Kuechler_2012, Steppke_2013a}

In the majority of modern dilatometers, samples are compressed between two anvils.\cite{Kuechler_2012} Therefore, a straightforward method to measure thermal expansion under uniaxial stress is to configure the dilatometer to also apply substantial force through these anvils.\cite{Kuechler_2016} It has been shown, however, that very high uniaxial stresses and high stress homogeneity can be achieved by preparing samples as narrow beams, embedding the ends in epoxy, and applying force through the epoxy and along the sample length.\cite{Barber_2019a} In this Letter, our goal is to explore methods for dilatometry measurements built around this sample configuration and to apply them to the hexagonal antiferromagnet Mn$_3$Sn. 

In Mn$_3$Sn, the Mn moments form a triangular spin structure below $T_N \approx \SI{420}{\kelvin}$.\cite{Tomiyoshi_1982a,Nagamiya_1982a} Recently, this chiral antiferromagnet has attracted considerable attention due to its anomalous transport properties\cite{Nakatsuji_2015a,Ikhlas_2017a} associated with the presence of magnetic Weyl fermions,\cite{Ikhlas_2017a,Kuroda_2017a} and the potential for applications in spintronics devices.\cite{Higo_2018a,Tsai_2020} However, stoichiometric Mn$_3$Sn undergoes a first-order transition below $T_{H}$ $\approx$ 280 K from the triangular spin structure to a spin spiral,\cite{Kren_1975, Cable_1993a} in which the topological transport properties are lost.\cite{Sung_2018,Song_2020} While room-temperature application of topological transport properties is therefore possible, this transition prevents their study at low temperature, where greater precision is available through the suppression of thermal fluctuations. By growing off-stoichiometry Mn$_{3+x}$Sn$_{1-x}$, this transition can be suppressed, but measurement precision then becomes limited by defects rather than thermal fluctuations. 

The transition at $T_H$ is accompanied by an expansion along [0001], and our goal here is to suppress $T_H$ as far as possible through compressive uniaxial stress, $\sigma$, which introduces no disorder when the sample deformation is elastic, applied along [0001]. From the Clausius-Clapeyron relationship, we estimate that $\sigma\approx -4$~GPa, where $\sigma<0$ denotes compression, would be required to obtain $T_H \rightarrow 0$. Here, we achieve $\sigma = -1.51$~GPa, and show through thermal expansion measurements that, at this stress, $T_H$ is suppressed by 54.6~K. 

A piezoelectric-driven uniaxial stress apparatus which incorporates both force and displacement sensors has been presented in Ref.\onlinecite{Barber_2019a}. Such a device could, in principle, be used for dilatometry measurements. In that device, the sample is coupled directly (i.e. through a high-stiffness mechanical link) to piezoelectric actuators. Force is thus delivered at a high spring constant, because piezoelectric actuators are, in general, high-stiffness devices that generate only small displacements. Consequently, the force applied to the sample is not automatically independent of length changes that the sample might undergo, and must instead be held constant through feedback. 

The approach we explore here is to place a spring of low spring constant between the piezoelectric actuators and the sample; we term this the conversion spring, as it converts displacement from the actuators into a force on the sample. The advantage is that force is delivered to the sample at a low spring constant. For materials such as Mn$_3$Sn that undergo first-order transitions, this means that the force on the sample does not change drastically even if the sample length changes abruptly. The disadvantage is, to achieve high stresses, the actuators must be long and/or the sample small. Here, the actuators can generate a displacement of $\sim$100 \SI{}{\micro\meter}. For the force on the sample to be well-controlled, the majority of this displacement must go into the conversion spring. If we specify that $\sim$90\% should go into the conversion spring, there is only $\sim$10 \SI{}{\micro\meter} left to compress the sample, as well as the epoxy that holds it and any other coupling elements to the sample.  Therefore to reach high strains we limit the sample length to $\sim$300 \SI{}{\micro\meter}.


\begin{figure}[!t]
	\centering
		\includegraphics[width=0.95\linewidth]{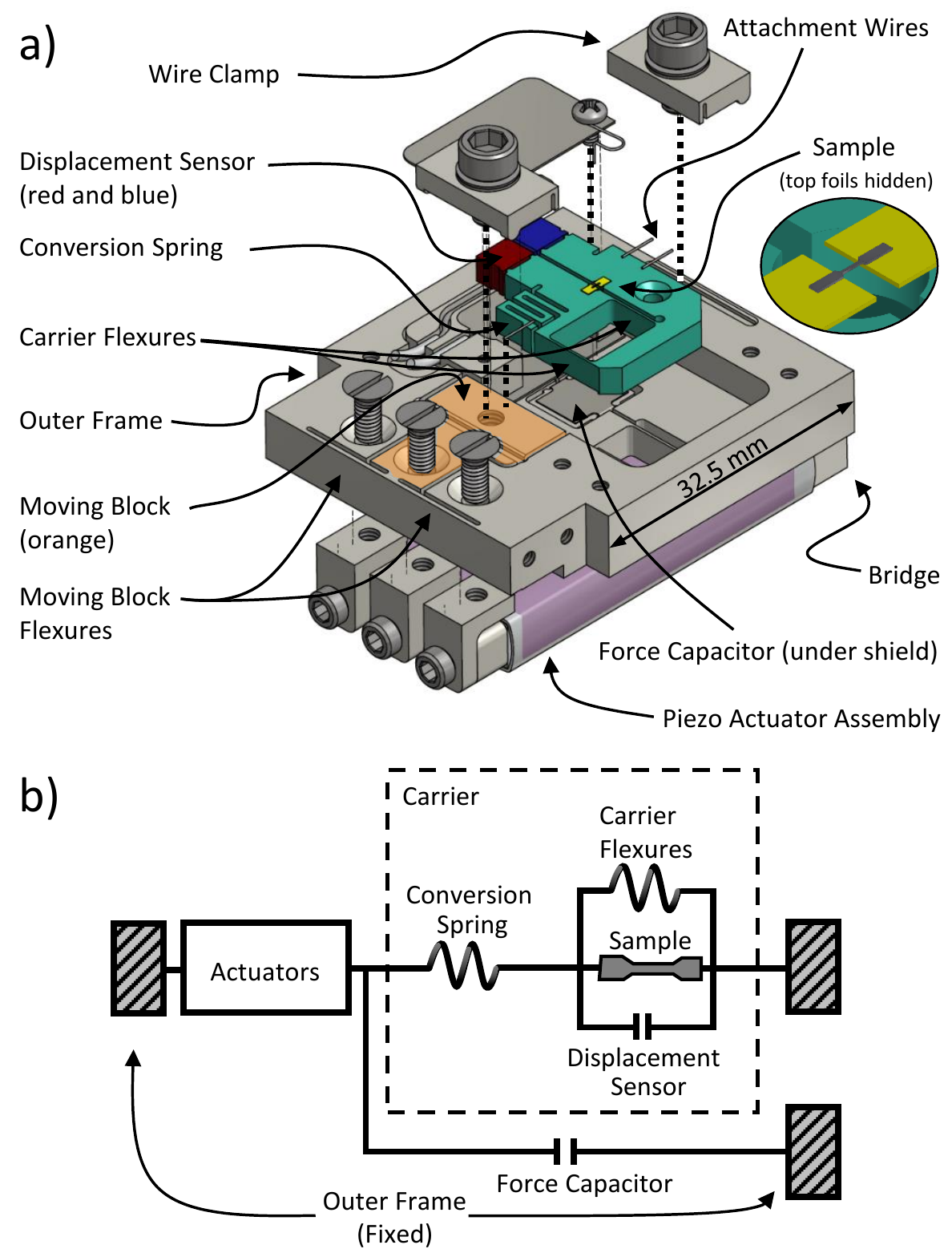}
		\caption[Drawings of the Stress Cell]
				{
				\label{fig:Stress_Cell}
\small				a) Illustration of the stress dilatometer. Piezoelectric actuators drive motion of a moving block, which is joined to the outer frame through flexures. To measure the change in sample length, the capacitance between the red and blue blocks of the displacement sensor is measured. b) A block diagram showing the essential mechanical connections. \vspace{-0.5cm}
				}
\end{figure}


\begin{figure}[!ht]
	\centering
		\includegraphics[width=\linewidth]{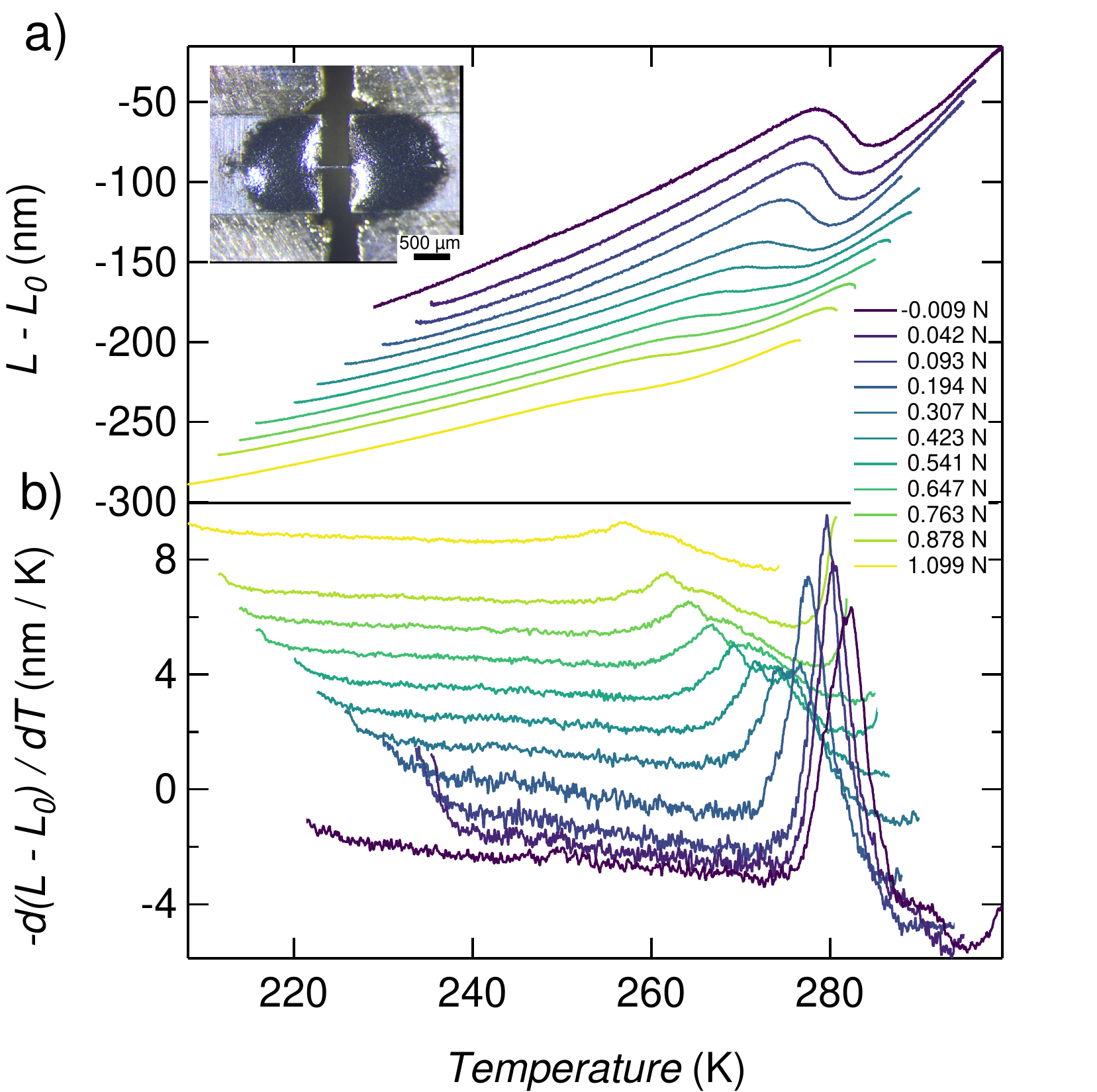}
		\caption[Displacement change vs temperature data]
				{
				\label{fig:BarData}
\small				a) Change in length of Sample 1 versus temperature under constant applied force. 1 N $= -0.93$ GPa at the center of the bar. Curves are offset for clarity. The inset shows an image of the sample.
b) Derivative of the displacement change versus temperature. The transition broadens appreciably with applied force. \vspace{-0.5cm}
				}
\end{figure}



 
Our design is presented in Fig. \ref{fig:Stress_Cell}. Parts, unless otherwise noted, are made out of Grade 2 titanium. The device is composed of a main body that contains an integrated moving block, guided by flexures. Movement of this block is driven by three piezoelectric actuators (PI PICMA\textsuperscript\textregistered \ P-885.91) that sit below the main body, are connected to the body by L-shaped brackets, and are joined at their opposite ends by a rectangular prism, a ``bridge''. Macor\textsuperscript\textregistered \ caps are fixed on both ends of the actuators for electrical isolation, important for high voltage operation in exchange gas. With the actuators fixed together in this manner, expansion of the outer actuator translates the moving block towards the sample carrier and compress the sample. Expansion of the inner actuator tension the sample. 

The sample is mounted onto a detachable carrier comprised of two parts, one moving and the other fixed, joined by flexures; the sample spans the moving and fixed parts. The carrier incorporates the conversion spring. It also incorporates a second capacitive sensor, comprised of two blocks separated by $\sim \SI{10}{\micro\meter}$, one attached to the moving part and the other to the fixed part. This sensor measures changes in length of the sample, and so is termed the displacement sensor. The rest of the carrier serves as part of the shielding for this capacitor. The carrier is coupled to the cell through attachment wires, here \SI{200}{\micro\meter}-diameter beryllium copper, which transmit longitudinal applied force to the sample while attenuating large inadvertent torques that may be applied when the carrier is clamped to the rest of the apparatus. A schematic of the essential mechanical connections of the apparatus is shown in Fig. \ref{fig:Stress_Cell}b).


\begin{figure*}[!ht]
	\centering
		\includegraphics[width=0.95\linewidth]{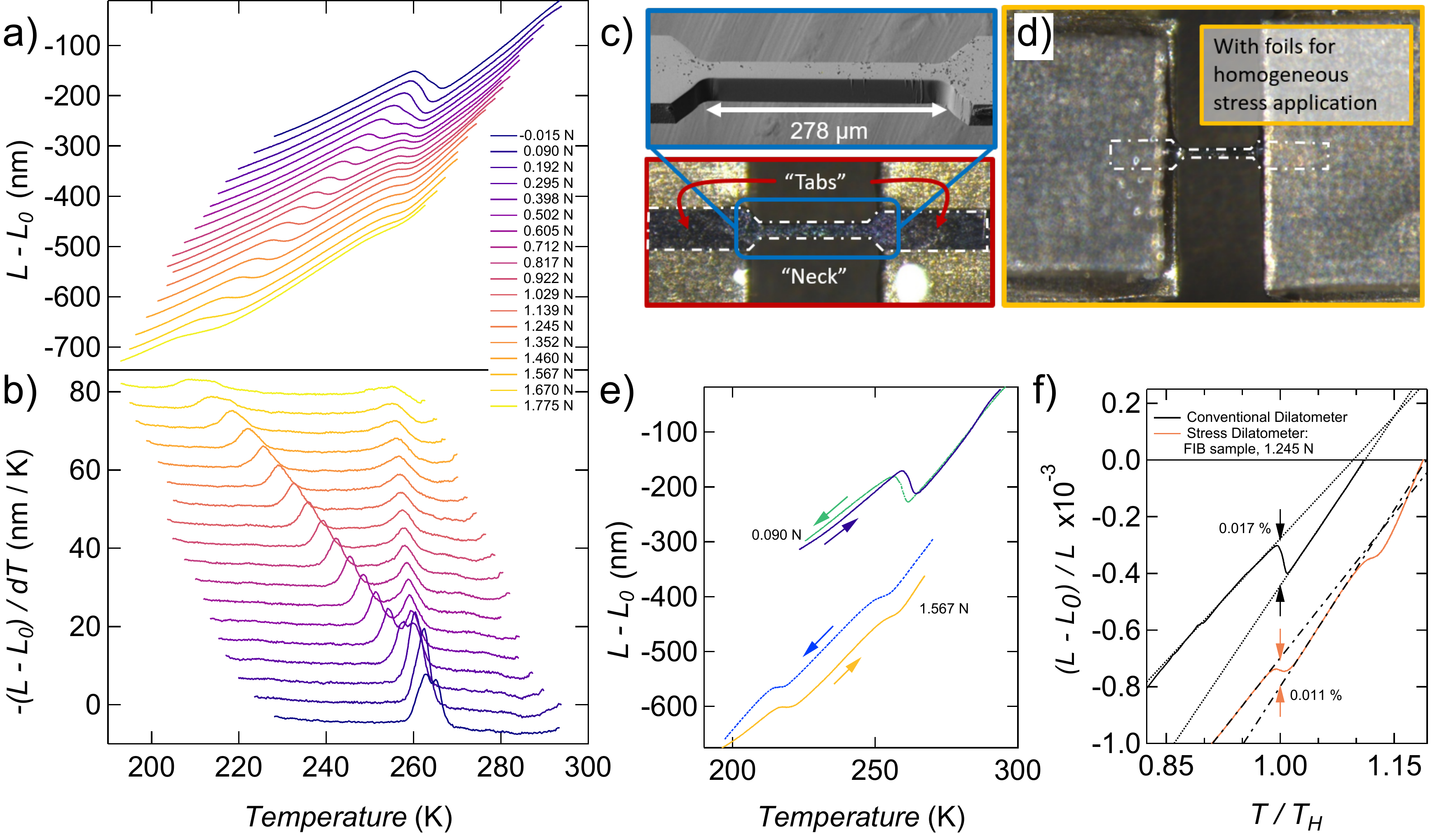}
		\caption[Displacement change vs temperature data]
				{
				\label{fig:data}
\small				a) Change in length of Sample 2 versus temperature under constant applied forces. Curves are offset for clarity. 1 N $= -0.87$ GPa in the sample neck. 
b) Derivative of the data in a) versus temperature. The presence of two transitions is due to the two distinct portions of the sample in contrast to Sample 1. By shaping the sample, the transitions stay narrow as stress increases.
c) Photograph and SEM micrograph of the Sample 2. 
d) Photograph of the sample after mounting, including the top Ti foils.
e) Hysteresis, as in a), for a low and high applied force.
f) Length changes of a third sample, measured near zero stress in a conventional dilatometer, and Sample 2, with $\sigma = -1.06$~GPa and $L = 278$~\SI{}{\micro\meter}. \vspace{-0.5cm}
}
\end{figure*}


Measured capacitances are converted to displacement, $d$, through $C = \varepsilon_0 A / (d+d_0) + C_\text{offset}$, where $A$ is the capacitor area and $C_\text{offset}$ is the stray capacitance. $A$ is 7.7 and 7.4 mm$^2$ for the force and displacement capacitors, respectively. $d_0$ is the initial capacitor plate separation, $\approx$150~$\mu$m for the force and $\approx$10~$\mu$m for the displacement capacitors. The stray capacitance was found to be negligible for both capacitors. The spring constants of the conversion spring and carrier flexures were measured at room temperature (see Supplementary Information) and are  $5.6\times10^4$ and $3.5\times10^4$ \SI{}{\newton\per\meter}, respectively. To obtain force, the length change measured by the displacement sensor was subtracted from that measured by the force sensor, to obtain the displacement applied to the conversion spring.

Single crystals of Mn$_3$Sn were grown from Sn flux using the Bridgman method, following protocols reported in a previous study. \cite{Ikhlas_2017a} After alignment using a Laue diffractometer, single crystals were spark eroded and polished into bar-like shapes and then screened through a combination of magnetometry and heat capacity measurements. The samples selected for study here have $280 > T_H > 260$~K, corresponding to Mn$_{3+x}$Sn$_{1-x}$ compositions with $0.009 < x < 0.02$, as measured by inductively coupled plasma spectroscopy.\footnote{M. Ikhlas and S. Nakatsuji, in preparation.} To provide a control measurement, the thermal expansion of a sample with $T_H = 267$~K was measured using a conventional capacitive dilatometer.\cite{Kuechler_2012}


Measurements were performed in an Oxford Instruments Teslatron\textsuperscript{TM}PT with a Cernox thermometer mounted to the outer frame of the stress dilatometer. An Andeen-Hagerling capacitance bridge, AH2550A, and a Keysight LCR meter, E4980AL, were used to measure the carrier and force capacitors respectively. To maintain a constant force sensor reading, the applied voltage to the actuators was driven with a proportional-integral-derivative controller. We find, with this experimental configuration, a noise level of 79 pm$/$$\sqrt{\SI{}{\hertz}}$ on the displacement sensor. The dominant contribution to this noise is fluctuations in the applied force from the feedback on the force capacitor. It could be improved by increasing the sensitivity of the force capacitor, e.g. by reducing its $d_0$, or lengthening the feedback time constant.

Samples are epoxied across a gap between two 50 $\mu$m-thick titanium foils. Sample 1, shown in Fig. \ref{fig:BarData}, was prepared with a uniform cross section of 30 $\SI{}{\micro\meter}$ x 36 $\SI{}{\micro\meter}$. $T_H$ is, as expected, suppressed by applied pressure, though the transition broadens dramatically as $|\sigma|$ is increased. We attribute this broadening to extended stress-gradient regions in the sample ends, which arise because force is transferred to the sample through the mounting epoxy over a substantial length scale, and the displacement sensor measures the total displacement across the carrier, including deformation of the epoxy and sample ends embedded in the epoxy.

Therefore, to concentrate stress efficiently, we found it essential to mill Sample 2 into a bow-tie shape, using  a Thermo Fisher Scientific Helios G4 PFIB UXe Xe$^{2+}$ plasma focused ion beam instrument, as shown in Fig.~\ref{fig:data}c). The two wide end portions of the sample are epoxied to the carrier, and serve to couple applied force from the carrier into the central, necked portion of the sample, with cross section of 32 $\SI{}{\micro\meter}$ x 36 $\SI{}{\micro\meter}$. By shaping the sample in this way, the shear stress within the epoxy layers is reduced, reducing the contribution to the measurement from the end portions. The sample is then epoxied across the gap between mounting foils, and additional foils are epoxied on top of the tabs to make the mounting symmetric, which improves stress homogeneity by reducing bending.\cite{Hicks_2014a}


\begin{figure}[t!]
        \centering 
				        \begin{subfigure}[b]{\linewidth}
                \centering
                \includegraphics[width=\linewidth]{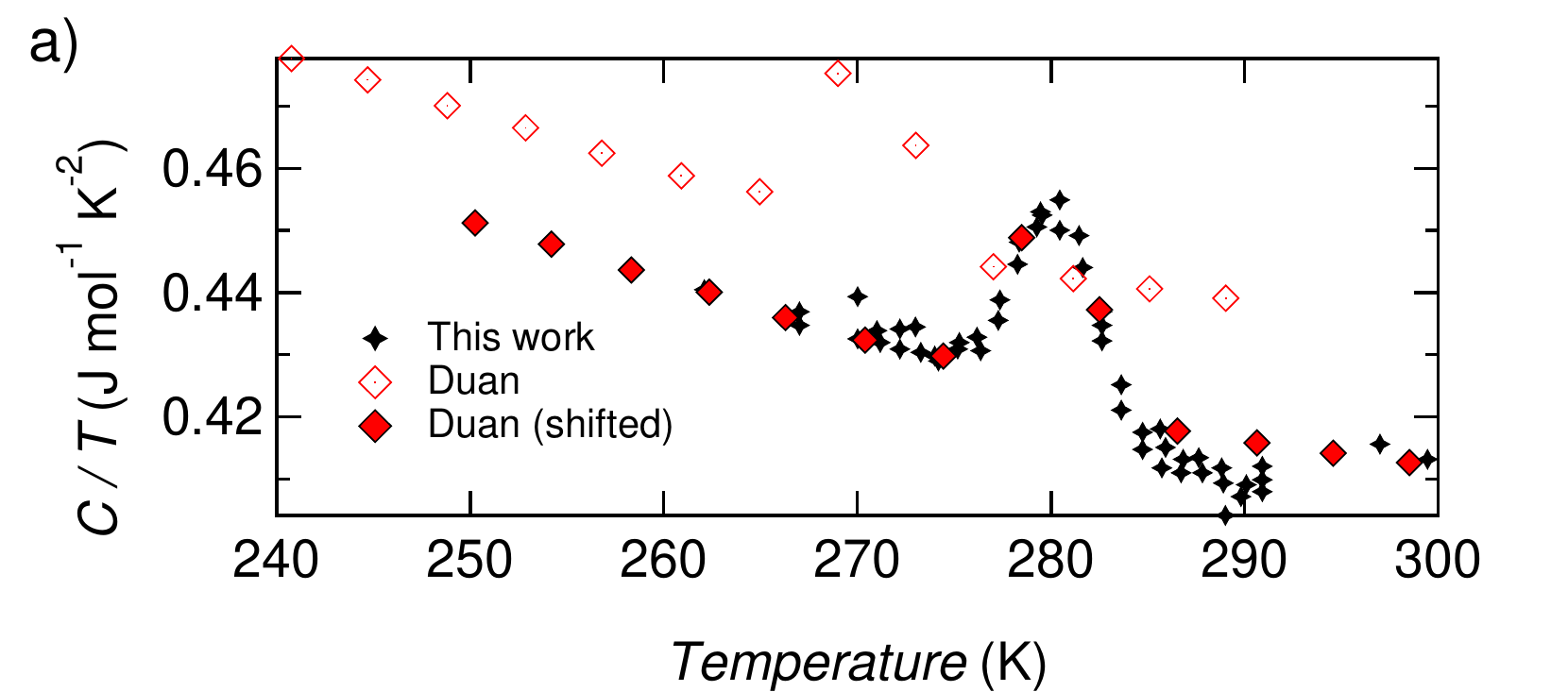}
                \caption*{}
                \label{subfig:}
        \end{subfigure}%
				
					\vspace{-0.75cm}
        \begin{subfigure}[b]{\linewidth}
                \centering
                \includegraphics[width=\linewidth]{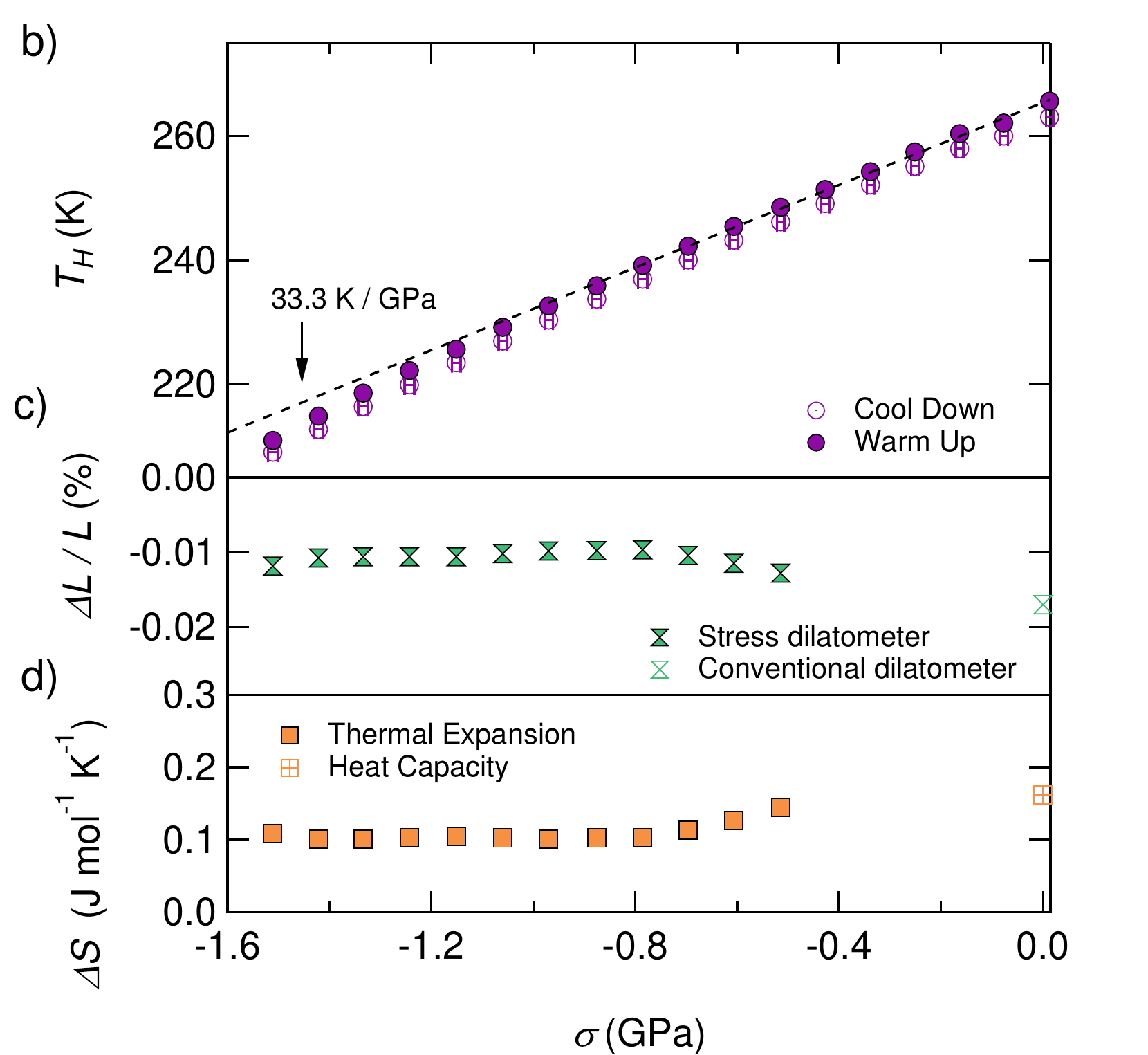}
                \caption*{}
                \label{subfig:}
        \end{subfigure}%
				\vspace{-0.5cm}
        \caption[Heat capacity and T vs P phase diagram]
						{
						\label{fig:phase}
\small						a) Heat capacity vs temperature in a Mn$_3$Sn sample with T$_H \approx$ 280 K(black). The red data are reproduced from Ref. \onlinecite{Duan_2015a}. b) AFM to spin spiral transition temperature vs uniaxial pressure in the FIB-prepared sample. Pressure values are inferred from the applied force and cross section of the neck region of the samples, and temperatures are taken from the peak of $-d\Delta L/dT$ in Fig.~\ref{fig:data}. c) Length change vs uniaxial pressure and entropy calculated with the Clausius-Clapeyron as explained in the text. Both quantities do not undergo a drastic response to the applied pressure within our experimental range. \vspace{-0.5cm}
						}
\end{figure}


Results are shown in Figs. \ref{fig:data} and \ref {fig:phase}. Fig. \ref{fig:data} panels a) and b) show the length change of Sample 2 and its derivative for forces from -0.02 to 1.78 N, equivalent to stresses of 0.01 to -1.51 GPa in the neck, applied along [0001]. Two transitions are apparent, one that remains at $\approx$260~K independent of the applied stress, and another that shifts to lower temperature as the sample is compressed. We attribute the stress-independent transition to the transition in the tabs of the sample, where the applied stress is low, and the transition that shifts to the transition in the necked portion of the sample. This is in contrast to the broadening seen in Sample 1, and shows that the shaping of the sample gives a sharp dichotomy between high- and low-stress sample regions. Thus, the contributions from the two sections can be cleanly separated. At -1.51 GPa, $T_H$ is suppressed by 54.6 K to 210.9 K. The width of the transition is nearly constant up to -1.33~GPa. Above $|\sigma|\sim$1.4 GPa the transition is broadened, which may be due to bending-induced strain inhomogeneity or plastic deformation prior to the sample fracture above $|\sigma|\sim$1.6 GPa.  Fig. \ref{fig:data}e) shows the hysteretic behavior vs. temperature. The first-order nature of the transition is shown by the step-like change in sample length, and hysteresis in $T_H$ between increasing- and decreasing-temperature ramps.

Fig. \ref{fig:phase}a) shows heat capacity data from a Mn$_3$Sn sample (black) with $T_H \approx 280$ K and a sample of reported composition Mn$_{3.03}$Sn$_{0.97}$ (red) with $T_H \approx 270$ K reproduced from Ref.~\onlinecite{Duan_2015a}. While the transition temperatures of the samples differ, the change in entropy at the transition appears to be similar. Fig. \ref{fig:phase}b) shows the dependence of $T_{H}$ on uniaxial pressure for Sample 2. The hysteresis between increasing-$T$ and decreasing-$T$ ramps is nearly stress-independent. Fig. \ref{fig:phase}c) shows the fractional length change of $\Delta L / L$ at $T_H$ for Sample 2 and that of an unstressed sample measured with a conventional dilatometer. Data points from Sample 2 are not included for pressures below $|\sigma|\sim$0.5 GPa, as the overlap between transitions in the neck and tab portions of the sample creates too much uncertainty. Our data in Fig. \ref{fig:phase}b) and c) can be related to the change in entropy at $T_H$, $\Delta S$, using the Clausius-Clapeyron relation:

\begin{equation}
- \left| dT_{H} / d\sigma\right| = (\Delta L/ L) V_{mol} / \Delta S
\label{eq:Clausius}
\end{equation}

which is valid for first order transitions. The negative sign on the left hand side of the equation is due to our convention that compression is negative. With this and our zero-pressure dilatometry and heat capacity data, we estimate the zero pressure $dT_{H} / d\sigma_{c}$ slope to be \SI{40.1}{\kelvin\per\giga\pascal}. $dT_H/d\sigma$ of Sample 2 is found to be 33.3 K GPa$^-1$ at low $|\sigma|$, in reasonable agreement with this estimate. The difference may reflect sample-to-sample variation. As $|\sigma|$ is increased to large values, $T_H$ deviates below this slope, see Fig. \ref{fig:phase}b). Applying the Clausius-Clapeyron relationship to the data of panel c), we find that the entropy change at the transition that the entropy change associated with the transition is mildly suppressed from its zero pressure value to \SI[mode=text]{0.1}{J.mol^{-1}.K^{-1}} above $|\sigma|\sim$0.7 \SI{}{\giga\pascal}, as shown in Fig. \ref{fig:phase}d).

From the estimated Young's modulus of Mn$_{3}$Sn along the $[0001]$ axis ($\sim$ 63 GPa, Supplementary Information), the sample strain at $\sigma = -1.51$~GPa is $\approx$2.4\%. As a comparison, it was found that 8\% substitution of the Sn site with Mn can suppress the spin spiral transition completely while changing the $c$ and $a$  lattice constants by only $\sim$ -0.3\%,\cite{Kren_1975} which indicates that the changes in $T_H$ due to Sn substitution are not primarily due to the effect of the substitution on the lattice constants. A similar case of composition sensitivity in a relatively high-temperature magnetic transition was observed in FeRh.\cite{Staunton_2014} In Mn$_3$Sn, it is striking that the spiral transition at $T_H$ is acutely sensitive to changes in composition, and yet, in comparison, mildly responds to lattice strains exceeding 2\%. This unexpected response to strain had to be directly measured; it could not be extrapolated from measurements at zero stress using the Clausius-Clapeyron relationship. Our results suggest that, with further technical refinement, suppression of $T_H$ to zero will be achievable, and the methodology presented here will be applicable to a range of other materials.


The authors thank Alexander Steppke and Hilary Noad for their technical expertise and fruitful scientific discussions. We also thank Thomas L{\"u}hmann for his help with LabVIEW programming. This work is partially supported by CREST(JPMJCR18T3 and JPMJCR15Q5), Japan Science and Technology Agency, by New Energy and Industrial Technology Development Organization (NEDO), by Grants-in-Aids for Scientific Research on Innovative Areas (15H05882 and 15H05883) from the Ministry of Education, Culture, Sports, Science, and Technology of Japan, and by Grants-in-Aid for Scientific Research (19H00650) from the Japanese Society for the Promotion of Science (JSPS). C.W.H. has 31\% ownership of Razorbill Instruments, a United Kingdom company that markets piezoelectric-based uniaxial stress cells.


The data that support the findings of this study are available within the article and its supplementary material. Raw data of this study are available from the corresponding author upon reasonable request.

\bibliography{2019.11_Mn3Sn_Stress_Dilatometry_Final}

\begin{thebibliography}{27}%
\makeatletter
\providecommand \@ifxundefined [1]{%
 \@ifx{#1\undefined}
}%
\providecommand \@ifnum [1]{%
 \ifnum #1\expandafter \@firstoftwo
 \else \expandafter \@secondoftwo
 \fi
}%
\providecommand \@ifx [1]{%
 \ifx #1\expandafter \@firstoftwo
 \else \expandafter \@secondoftwo
 \fi
}%
\providecommand \natexlab [1]{#1}%
\providecommand \enquote  [1]{``#1''}%
\providecommand \bibnamefont  [1]{#1}%
\providecommand \bibfnamefont [1]{#1}%
\providecommand \citenamefont [1]{#1}%
\providecommand \href@noop [0]{\@secondoftwo}%
\providecommand \href [0]{\begingroup \@sanitize@url \@href}%
\providecommand \@href[1]{\@@startlink{#1}\@@href}%
\providecommand \@@href[1]{\endgroup#1\@@endlink}%
\providecommand \@sanitize@url [0]{\catcode `\\12\catcode `\$12\catcode
  `\&12\catcode `\#12\catcode `\^12\catcode `\_12\catcode `\%12\relax}%
\providecommand \@@startlink[1]{}%
\providecommand \@@endlink[0]{}%
\providecommand \url  [0]{\begingroup\@sanitize@url \@url }%
\providecommand \@url [1]{\endgroup\@href {#1}{\urlprefix }}%
\providecommand \urlprefix  [0]{URL }%
\providecommand \Eprint [0]{\href }%
\providecommand \doibase [0]{http://dx.doi.org/}%
\providecommand \selectlanguage [0]{\@gobble}%
\providecommand \bibinfo  [0]{\@secondoftwo}%
\providecommand \bibfield  [0]{\@secondoftwo}%
\providecommand \translation [1]{[#1]}%
\providecommand \BibitemOpen [0]{}%
\providecommand \bibitemStop [0]{}%
\providecommand \bibitemNoStop [0]{.\EOS\space}%
\providecommand \EOS [0]{\spacefactor3000\relax}%
\providecommand \BibitemShut  [1]{\csname bibitem#1\endcsname}%
\let\auto@bib@innerbib\@empty
\bibitem [{\citenamefont {Kim}\ \emph {et~al.}(2018)\citenamefont {Kim},
  \citenamefont {Souliou}, \citenamefont {Barber}, \citenamefont
  {Lefran{\c{c}}ois}, \citenamefont {Minola}, \citenamefont {Tortora},
  \citenamefont {Heid}, \citenamefont {Nandi}, \citenamefont {Borzi},
  \citenamefont {Garbarino}, \citenamefont {Bosak}, \citenamefont {Porras},
  \citenamefont {Loew}, \citenamefont {König}, \citenamefont {Moll},
  \citenamefont {Mackenzie}, \citenamefont {Keimer}, \citenamefont {Hicks},\
  and\ \citenamefont {Tacon}}]{Kim_2018a}%
  \BibitemOpen
  \bibfield  {author} {\bibinfo {author} {\bibfnamefont {H.-H.}\ \bibnamefont
  {Kim}}, \bibinfo {author} {\bibfnamefont {S.~M.}\ \bibnamefont {Souliou}},
  \bibinfo {author} {\bibfnamefont {M.~E.}\ \bibnamefont {Barber}}, \bibinfo
  {author} {\bibfnamefont {E.}~\bibnamefont {Lefran{\c{c}}ois}}, \bibinfo
  {author} {\bibfnamefont {M.}~\bibnamefont {Minola}}, \bibinfo {author}
  {\bibfnamefont {M.}~\bibnamefont {Tortora}}, \bibinfo {author} {\bibfnamefont
  {R.}~\bibnamefont {Heid}}, \bibinfo {author} {\bibfnamefont {N.}~\bibnamefont
  {Nandi}}, \bibinfo {author} {\bibfnamefont {R.~A.}\ \bibnamefont {Borzi}},
  \bibinfo {author} {\bibfnamefont {G.}~\bibnamefont {Garbarino}}, \bibinfo
  {author} {\bibfnamefont {A.}~\bibnamefont {Bosak}}, \bibinfo {author}
  {\bibfnamefont {J.}~\bibnamefont {Porras}}, \bibinfo {author} {\bibfnamefont
  {T.}~\bibnamefont {Loew}}, \bibinfo {author} {\bibfnamefont {M.}~\bibnamefont
  {König}}, \bibinfo {author} {\bibfnamefont {P.~J.~W.}\ \bibnamefont {Moll}},
  \bibinfo {author} {\bibfnamefont {A.~P.}\ \bibnamefont {Mackenzie}}, \bibinfo
  {author} {\bibfnamefont {B.}~\bibnamefont {Keimer}}, \bibinfo {author}
  {\bibfnamefont {C.~W.}\ \bibnamefont {Hicks}}, \ and\ \bibinfo {author}
  {\bibfnamefont {M.~L.}\ \bibnamefont {Tacon}},\ }\bibfield  {title} {\enquote
  {\bibinfo {title} {Uniaxial pressure control of competing orders in a
  high-temperature superconductor},}\ }\href {\doibase 10.1126/science.aat4708}
  {\bibfield  {journal} {\bibinfo  {journal} {Science}\ }\textbf {\bibinfo
  {volume} {362}},\ \bibinfo {pages} {1040--1044} (\bibinfo {year}
  {2018})}\BibitemShut {NoStop}%
\bibitem [{\citenamefont {Steppke}\ \emph {et~al.}(2017)\citenamefont
  {Steppke}, \citenamefont {Zhao}, \citenamefont {Barber}, \citenamefont
  {Scaffidi}, \citenamefont {Jerzembeck}, \citenamefont {Rosner}, \citenamefont
  {Gibbs}, \citenamefont {Maeno}, \citenamefont {Simon}, \citenamefont
  {Mackenzie},\ and\ \citenamefont {Hicks}}]{Steppke_2017a}%
  \BibitemOpen
  \bibfield  {author} {\bibinfo {author} {\bibfnamefont {A.}~\bibnamefont
  {Steppke}}, \bibinfo {author} {\bibfnamefont {L.}~\bibnamefont {Zhao}},
  \bibinfo {author} {\bibfnamefont {M.~E.}\ \bibnamefont {Barber}}, \bibinfo
  {author} {\bibfnamefont {T.}~\bibnamefont {Scaffidi}}, \bibinfo {author}
  {\bibfnamefont {F.}~\bibnamefont {Jerzembeck}}, \bibinfo {author}
  {\bibfnamefont {H.}~\bibnamefont {Rosner}}, \bibinfo {author} {\bibfnamefont
  {A.~S.}\ \bibnamefont {Gibbs}}, \bibinfo {author} {\bibfnamefont
  {Y.}~\bibnamefont {Maeno}}, \bibinfo {author} {\bibfnamefont {S.~H.}\
  \bibnamefont {Simon}}, \bibinfo {author} {\bibfnamefont {A.~P.}\ \bibnamefont
  {Mackenzie}}, \ and\ \bibinfo {author} {\bibfnamefont {C.~W.}\ \bibnamefont
  {Hicks}},\ }\bibfield  {title} {\enquote {\bibinfo {title} {{Strong peak in
  T$_c$ of Sr$_2$RuO$_4$ under uniaxial pressure}},}\ }\href {\doibase
  10.1126/science.aaf9398} {\bibfield  {journal} {\bibinfo  {journal}
  {Science}\ }\textbf {\bibinfo {volume} {355}},\ \bibinfo {pages} {eaaf9398}
  (\bibinfo {year} {2017})}\BibitemShut {NoStop}%
\bibitem [{\citenamefont {Barber}\ \emph {et~al.}(2018)\citenamefont {Barber},
  \citenamefont {Gibbs}, \citenamefont {Maeno}, \citenamefont {Mackenzie},\
  and\ \citenamefont {Hicks}}]{Barber_2018a}%
  \BibitemOpen
  \bibfield  {author} {\bibinfo {author} {\bibfnamefont {M.}~\bibnamefont
  {Barber}}, \bibinfo {author} {\bibfnamefont {A.}~\bibnamefont {Gibbs}},
  \bibinfo {author} {\bibfnamefont {Y.}~\bibnamefont {Maeno}}, \bibinfo
  {author} {\bibfnamefont {A.}~\bibnamefont {Mackenzie}}, \ and\ \bibinfo
  {author} {\bibfnamefont {C.}~\bibnamefont {Hicks}},\ }\bibfield  {title}
  {\enquote {\bibinfo {title} {{Resistivity in the Vicinity of a van Hove
  Singularity}: {Sr}$_2${RuO}$_4$ {under Uniaxial Pressure}},}\ }\href
  {\doibase 10.1103/physrevlett.120.076602} {\bibfield  {journal} {\bibinfo
  {journal} {Physical Review Letters}\ }\textbf {\bibinfo {volume} {120}}
  (\bibinfo {year} {2018}),\ 10.1103/physrevlett.120.076602}\BibitemShut
  {NoStop}%
\bibitem [{\citenamefont {Kissikov}\ \emph {et~al.}(2018)\citenamefont
  {Kissikov}, \citenamefont {Sarkar}, \citenamefont {Lawson}, \citenamefont
  {Bush}, \citenamefont {Timmons}, \citenamefont {Tanatar}, \citenamefont
  {Prozorov}, \citenamefont {Bud'ko}, \citenamefont {Canfield}, \citenamefont
  {Fernandes},\ and\ \citenamefont {Curro}}]{Kissikov_2018a}%
  \BibitemOpen
  \bibfield  {author} {\bibinfo {author} {\bibfnamefont {T.}~\bibnamefont
  {Kissikov}}, \bibinfo {author} {\bibfnamefont {R.}~\bibnamefont {Sarkar}},
  \bibinfo {author} {\bibfnamefont {M.}~\bibnamefont {Lawson}}, \bibinfo
  {author} {\bibfnamefont {B.~T.}\ \bibnamefont {Bush}}, \bibinfo {author}
  {\bibfnamefont {E.~I.}\ \bibnamefont {Timmons}}, \bibinfo {author}
  {\bibfnamefont {M.~A.}\ \bibnamefont {Tanatar}}, \bibinfo {author}
  {\bibfnamefont {R.}~\bibnamefont {Prozorov}}, \bibinfo {author}
  {\bibfnamefont {S.~L.}\ \bibnamefont {Bud'ko}}, \bibinfo {author}
  {\bibfnamefont {P.~C.}\ \bibnamefont {Canfield}}, \bibinfo {author}
  {\bibfnamefont {R.~M.}\ \bibnamefont {Fernandes}}, \ and\ \bibinfo {author}
  {\bibfnamefont {N.~J.}\ \bibnamefont {Curro}},\ }\bibfield  {title} {\enquote
  {\bibinfo {title} {Uniaxial strain control of spin-polarization in
  multicomponent nematic order of {BaFe}$_2$as$_2$},}\ }\href {\doibase
  10.1038/s41467-018-03377-8} {\bibfield  {journal} {\bibinfo  {journal}
  {Nature Communications}\ }\textbf {\bibinfo {volume} {9}} (\bibinfo {year}
  {2018}),\ 10.1038/s41467-018-03377-8}\BibitemShut {NoStop}%
\bibitem [{\citenamefont {Grinenko}\ \emph {et~al.}()\citenamefont {Grinenko},
  \citenamefont {Ghosh}, \citenamefont {Sarkar}, \citenamefont {Orain},
  \citenamefont {Nikitin}, \citenamefont {Elender}, \citenamefont {Das},
  \citenamefont {Guguchia}, \citenamefont {Brückner}, \citenamefont {Barber},
  \citenamefont {Park}, \citenamefont {Kikugawa}, \citenamefont {Sokolov},
  \citenamefont {Bobowski}, \citenamefont {Miyoshi}, \citenamefont {Maeno},
  \citenamefont {Mackenzie}, \citenamefont {Luetkens}, \citenamefont {Hicks},\
  and\ \citenamefont {Klauss}}]{Grinenko_2020a}%
  \BibitemOpen
  \bibfield  {author} {\bibinfo {author} {\bibfnamefont {V.}~\bibnamefont
  {Grinenko}}, \bibinfo {author} {\bibfnamefont {S.}~\bibnamefont {Ghosh}},
  \bibinfo {author} {\bibfnamefont {R.}~\bibnamefont {Sarkar}}, \bibinfo
  {author} {\bibfnamefont {J.-C.}\ \bibnamefont {Orain}}, \bibinfo {author}
  {\bibfnamefont {A.}~\bibnamefont {Nikitin}}, \bibinfo {author} {\bibfnamefont
  {M.}~\bibnamefont {Elender}}, \bibinfo {author} {\bibfnamefont
  {D.}~\bibnamefont {Das}}, \bibinfo {author} {\bibfnamefont {Z.}~\bibnamefont
  {Guguchia}}, \bibinfo {author} {\bibfnamefont {F.}~\bibnamefont {Brückner}},
  \bibinfo {author} {\bibfnamefont {M.~E.}\ \bibnamefont {Barber}}, \bibinfo
  {author} {\bibfnamefont {J.}~\bibnamefont {Park}}, \bibinfo {author}
  {\bibfnamefont {N.}~\bibnamefont {Kikugawa}}, \bibinfo {author}
  {\bibfnamefont {D.~A.}\ \bibnamefont {Sokolov}}, \bibinfo {author}
  {\bibfnamefont {J.~S.}\ \bibnamefont {Bobowski}}, \bibinfo {author}
  {\bibfnamefont {T.}~\bibnamefont {Miyoshi}}, \bibinfo {author} {\bibfnamefont
  {Y.}~\bibnamefont {Maeno}}, \bibinfo {author} {\bibfnamefont {A.~P.}\
  \bibnamefont {Mackenzie}}, \bibinfo {author} {\bibfnamefont {H.}~\bibnamefont
  {Luetkens}}, \bibinfo {author} {\bibfnamefont {C.~W.}\ \bibnamefont {Hicks}},
  \ and\ \bibinfo {author} {\bibfnamefont {H.-H.}\ \bibnamefont {Klauss}},\
  }\bibfield  {title} {\enquote {\bibinfo {title} {Split superconducting and
  time-reversal symmetry-breaking transitions, and magnetic order in
  sr$_2$ruo$_4$ under uniaxial stress},}\ }\href@noop {} {\ }\Eprint
  {http://arxiv.org/abs/2001.08152v2} {2001.08152v2} \BibitemShut {NoStop}%
\bibitem [{\citenamefont {Meingast}\ \emph {et~al.}(1990)\citenamefont
  {Meingast}, \citenamefont {Blank}, \citenamefont {B{\"u}rkle}, \citenamefont
  {Obst}, \citenamefont {Wolf}, \citenamefont {Wühl}, \citenamefont
  {Selvamanickam},\ and\ \citenamefont {Salama}}]{Meingast_1990a}%
  \BibitemOpen
  \bibfield  {author} {\bibinfo {author} {\bibfnamefont {C.}~\bibnamefont
  {Meingast}}, \bibinfo {author} {\bibfnamefont {B.}~\bibnamefont {Blank}},
  \bibinfo {author} {\bibfnamefont {H.}~\bibnamefont {B{\"u}rkle}}, \bibinfo
  {author} {\bibfnamefont {B.}~\bibnamefont {Obst}}, \bibinfo {author}
  {\bibfnamefont {T.}~\bibnamefont {Wolf}}, \bibinfo {author} {\bibfnamefont
  {H.}~\bibnamefont {Wühl}}, \bibinfo {author} {\bibfnamefont
  {V.}~\bibnamefont {Selvamanickam}}, \ and\ \bibinfo {author} {\bibfnamefont
  {K.}~\bibnamefont {Salama}},\ }\bibfield  {title} {\enquote {\bibinfo {title}
  {Anisotropic pressure dependence of $t_{c}$ in single-crystal
  {YBa}$_2${Cu}$_3${O}$_7$ via thermal expansion},}\ }\href {\doibase
  10.1103/physrevb.41.11299} {\bibfield  {journal} {\bibinfo  {journal}
  {Physical Review B}\ }\textbf {\bibinfo {volume} {41}},\ \bibinfo {pages}
  {11299--11304} (\bibinfo {year} {1990})}\BibitemShut {NoStop}%
\bibitem [{\citenamefont {Meingast}\ \emph {et~al.}(1991)\citenamefont
  {Meingast}, \citenamefont {Kraut}, \citenamefont {Wolf}, \citenamefont
  {Wühl}, \citenamefont {Erb},\ and\ \citenamefont
  {Müller-Vogt}}]{Meingast_1991a}%
  \BibitemOpen
  \bibfield  {author} {\bibinfo {author} {\bibfnamefont {C.}~\bibnamefont
  {Meingast}}, \bibinfo {author} {\bibfnamefont {O.}~\bibnamefont {Kraut}},
  \bibinfo {author} {\bibfnamefont {T.}~\bibnamefont {Wolf}}, \bibinfo {author}
  {\bibfnamefont {H.}~\bibnamefont {Wühl}}, \bibinfo {author} {\bibfnamefont
  {A.}~\bibnamefont {Erb}}, \ and\ \bibinfo {author} {\bibfnamefont
  {G.}~\bibnamefont {Müller-Vogt}},\ }\bibfield  {title} {\enquote {\bibinfo
  {title} {Large ${a-b}$ anisotropy of the expansivity anomaly at $t_{c}$ in
  untwinned {YBa}$_2$cu$_3$o$_{7-\delta}$},}\ }\href {\doibase
  10.1103/physrevlett.67.1634} {\bibfield  {journal} {\bibinfo  {journal}
  {Physical Review Letters}\ }\textbf {\bibinfo {volume} {67}},\ \bibinfo
  {pages} {1634--1637} (\bibinfo {year} {1991})}\BibitemShut {NoStop}%
\bibitem [{\citenamefont {Westerkamp}\ \emph {et~al.}(2009)\citenamefont
  {Westerkamp}, \citenamefont {Deppe}, \citenamefont {K{\"u}chler},
  \citenamefont {Brando}, \citenamefont {Geibel}, \citenamefont {Gegenwart},
  \citenamefont {Pikul},\ and\ \citenamefont {Steglich}}]{Westerkamp_2009a}%
  \BibitemOpen
  \bibfield  {author} {\bibinfo {author} {\bibfnamefont {T.}~\bibnamefont
  {Westerkamp}}, \bibinfo {author} {\bibfnamefont {M.}~\bibnamefont {Deppe}},
  \bibinfo {author} {\bibfnamefont {R.}~\bibnamefont {K{\"u}chler}}, \bibinfo
  {author} {\bibfnamefont {M.}~\bibnamefont {Brando}}, \bibinfo {author}
  {\bibfnamefont {C.}~\bibnamefont {Geibel}}, \bibinfo {author} {\bibfnamefont
  {P.}~\bibnamefont {Gegenwart}}, \bibinfo {author} {\bibfnamefont {A.~P.}\
  \bibnamefont {Pikul}}, \ and\ \bibinfo {author} {\bibfnamefont
  {F.}~\bibnamefont {Steglich}},\ }\bibfield  {title} {\enquote {\bibinfo
  {title} {{Kondo-Cluster-Glass State near a Ferromagnetic Quantum Phase
  Transition}},}\ }\href {\doibase 10.1103/physrevlett.102.206404} {\bibfield
  {journal} {\bibinfo  {journal} {Physical Review Letters}\ }\textbf {\bibinfo
  {volume} {102}} (\bibinfo {year} {2009}),\
  10.1103/physrevlett.102.206404}\BibitemShut {NoStop}%
\bibitem [{\citenamefont {K{\"u}chler}\ \emph {et~al.}(2012)\citenamefont
  {K{\"u}chler}, \citenamefont {Bauer}, \citenamefont {Brando},\ and\
  \citenamefont {Steglich}}]{Kuechler_2012}%
  \BibitemOpen
  \bibfield  {author} {\bibinfo {author} {\bibfnamefont {R.}~\bibnamefont
  {K{\"u}chler}}, \bibinfo {author} {\bibfnamefont {T.}~\bibnamefont {Bauer}},
  \bibinfo {author} {\bibfnamefont {M.}~\bibnamefont {Brando}}, \ and\ \bibinfo
  {author} {\bibfnamefont {F.}~\bibnamefont {Steglich}},\ }\bibfield  {title}
  {\enquote {\bibinfo {title} {{A compact and miniaturized high resolution
  capacitance dilatometer for measuring thermal expansion and
  magnetostriction}},}\ }\href {\doibase 10.1063/1.4748864} {\bibfield
  {journal} {\bibinfo  {journal} {Review of Scientific Instruments}\ }\textbf
  {\bibinfo {volume} {83}},\ \bibinfo {pages} {095102} (\bibinfo {year}
  {2012})}\BibitemShut {NoStop}%
\bibitem [{\citenamefont {Steppke}\ \emph {et~al.}(2013)\citenamefont
  {Steppke}, \citenamefont {K{\"u}chler}, \citenamefont {Lausberg},
  \citenamefont {Lengyel}, \citenamefont {Steinke}, \citenamefont {Borth},
  \citenamefont {Luhmann}, \citenamefont {Krellner}, \citenamefont {Nicklas},
  \citenamefont {Geibel}, \citenamefont {Steglich},\ and\ \citenamefont
  {Brando}}]{Steppke_2013a}%
  \BibitemOpen
  \bibfield  {author} {\bibinfo {author} {\bibfnamefont {A.}~\bibnamefont
  {Steppke}}, \bibinfo {author} {\bibfnamefont {R.}~\bibnamefont
  {K{\"u}chler}}, \bibinfo {author} {\bibfnamefont {S.}~\bibnamefont
  {Lausberg}}, \bibinfo {author} {\bibfnamefont {E.}~\bibnamefont {Lengyel}},
  \bibinfo {author} {\bibfnamefont {L.}~\bibnamefont {Steinke}}, \bibinfo
  {author} {\bibfnamefont {R.}~\bibnamefont {Borth}}, \bibinfo {author}
  {\bibfnamefont {T.}~\bibnamefont {Luhmann}}, \bibinfo {author} {\bibfnamefont
  {C.}~\bibnamefont {Krellner}}, \bibinfo {author} {\bibfnamefont
  {M.}~\bibnamefont {Nicklas}}, \bibinfo {author} {\bibfnamefont
  {C.}~\bibnamefont {Geibel}}, \bibinfo {author} {\bibfnamefont
  {F.}~\bibnamefont {Steglich}}, \ and\ \bibinfo {author} {\bibfnamefont
  {M.}~\bibnamefont {Brando}},\ }\bibfield  {title} {\enquote {\bibinfo {title}
  {{Ferromagnetic Quantum Critical Point in the Heavy-Fermion Metal}
  {YbNi}$_4$({P}$_{1-x}${As}$_x$)$_2$},}\ }\href {\doibase
  10.1126/science.1230583} {\bibfield  {journal} {\bibinfo  {journal}
  {Science}\ }\textbf {\bibinfo {volume} {339}},\ \bibinfo {pages} {933--936}
  (\bibinfo {year} {2013})}\BibitemShut {NoStop}%
\bibitem [{\citenamefont {K{\"u}chler}, \citenamefont {Stingl},\ and\
  \citenamefont {Gegenwart}(2016)}]{Kuechler_2016}%
  \BibitemOpen
  \bibfield  {author} {\bibinfo {author} {\bibfnamefont {R.}~\bibnamefont
  {K{\"u}chler}}, \bibinfo {author} {\bibfnamefont {C.}~\bibnamefont {Stingl}},
  \ and\ \bibinfo {author} {\bibfnamefont {P.}~\bibnamefont {Gegenwart}},\
  }\bibfield  {title} {\enquote {\bibinfo {title} {{A uniaxial stress
  capacitive dilatometer for high-resolution thermal expansion and
  magnetostriction under multiextreme conditions}},}\ }\href {\doibase
  10.1063/1.4958957} {\bibfield  {journal} {\bibinfo  {journal} {Review of
  Scientific Instruments}\ }\textbf {\bibinfo {volume} {87}},\ \bibinfo {pages}
  {073903} (\bibinfo {year} {2016})}\BibitemShut {NoStop}%
\bibitem [{\citenamefont {Barber}\ \emph {et~al.}(2019)\citenamefont {Barber},
  \citenamefont {Steppke}, \citenamefont {Mackenzie},\ and\ \citenamefont
  {Hicks}}]{Barber_2019a}%
  \BibitemOpen
  \bibfield  {author} {\bibinfo {author} {\bibfnamefont {M.~E.}\ \bibnamefont
  {Barber}}, \bibinfo {author} {\bibfnamefont {A.}~\bibnamefont {Steppke}},
  \bibinfo {author} {\bibfnamefont {A.~P.}\ \bibnamefont {Mackenzie}}, \ and\
  \bibinfo {author} {\bibfnamefont {C.~W.}\ \bibnamefont {Hicks}},\ }\bibfield
  {title} {\enquote {\bibinfo {title} {{Piezoelectric-based uniaxial pressure
  cell with integrated force and displacement sensors}},}\ }\href {\doibase
  10.1063/1.5075485} {\bibfield  {journal} {\bibinfo  {journal} {Review of
  Scientific Instruments}\ }\textbf {\bibinfo {volume} {90}},\ \bibinfo {pages}
  {023904} (\bibinfo {year} {2019})}\BibitemShut {NoStop}%
\bibitem [{\citenamefont {Tomiyoshi}\ and\ \citenamefont
  {Yamaguchi}(1982)}]{Tomiyoshi_1982a}%
  \BibitemOpen
  \bibfield  {author} {\bibinfo {author} {\bibfnamefont {S.}~\bibnamefont
  {Tomiyoshi}}\ and\ \bibinfo {author} {\bibfnamefont {Y.}~\bibnamefont
  {Yamaguchi}},\ }\bibfield  {title} {\enquote {\bibinfo {title} {{Magnetic
  Structure and Weak Ferromagnetism of Mn$_3$Sn Studied by Polarized Neutron
  Diffraction}},}\ }\href {\doibase 10.1143/jpsj.51.2478} {\bibfield  {journal}
  {\bibinfo  {journal} {Journal of the Physical Society of Japan}\ }\textbf
  {\bibinfo {volume} {51}},\ \bibinfo {pages} {2478--2486} (\bibinfo {year}
  {1982})}\BibitemShut {NoStop}%
\bibitem [{\citenamefont {Nagamiya}, \citenamefont {Tomiyoshi},\ and\
  \citenamefont {Yamaguchi}(1982)}]{Nagamiya_1982a}%
  \BibitemOpen
  \bibfield  {author} {\bibinfo {author} {\bibfnamefont {T.}~\bibnamefont
  {Nagamiya}}, \bibinfo {author} {\bibfnamefont {S.}~\bibnamefont {Tomiyoshi}},
  \ and\ \bibinfo {author} {\bibfnamefont {Y.}~\bibnamefont {Yamaguchi}},\
  }\bibfield  {title} {\enquote {\bibinfo {title} {{Triangular spin
  configuration and weak ferromagnetism of Mn$_3$Sn and Mn$_3$Ge}},}\ }\href
  {\doibase 10.1016/0038-1098(82)90159-4} {\bibfield  {journal} {\bibinfo
  {journal} {Solid State Communications}\ }\textbf {\bibinfo {volume} {42}},\
  \bibinfo {pages} {385--388} (\bibinfo {year} {1982})}\BibitemShut {NoStop}%
\bibitem [{\citenamefont {Nakatsuji}, \citenamefont {Kiyohara},\ and\
  \citenamefont {Higo}(2015)}]{Nakatsuji_2015a}%
  \BibitemOpen
  \bibfield  {author} {\bibinfo {author} {\bibfnamefont {S.}~\bibnamefont
  {Nakatsuji}}, \bibinfo {author} {\bibfnamefont {N.}~\bibnamefont {Kiyohara}},
  \ and\ \bibinfo {author} {\bibfnamefont {T.}~\bibnamefont {Higo}},\
  }\bibfield  {title} {\enquote {\bibinfo {title} {{Large anomalous Hall effect
  in a non-collinear antiferromagnet at room temperature}},}\ }\href {\doibase
  10.1038/nature15723} {\bibfield  {journal} {\bibinfo  {journal} {Nature}\
  }\textbf {\bibinfo {volume} {527}},\ \bibinfo {pages} {212--215} (\bibinfo
  {year} {2015})}\BibitemShut {NoStop}%
\bibitem [{\citenamefont {Ikhlas}\ \emph {et~al.}(2017)\citenamefont {Ikhlas},
  \citenamefont {Tomita}, \citenamefont {Koretsune}, \citenamefont {Suzuki},
  \citenamefont {Nishio-Hamane}, \citenamefont {Arita}, \citenamefont {Otani},\
  and\ \citenamefont {Nakatsuji}}]{Ikhlas_2017a}%
  \BibitemOpen
  \bibfield  {author} {\bibinfo {author} {\bibfnamefont {M.}~\bibnamefont
  {Ikhlas}}, \bibinfo {author} {\bibfnamefont {T.}~\bibnamefont {Tomita}},
  \bibinfo {author} {\bibfnamefont {T.}~\bibnamefont {Koretsune}}, \bibinfo
  {author} {\bibfnamefont {M.-T.}\ \bibnamefont {Suzuki}}, \bibinfo {author}
  {\bibfnamefont {D.}~\bibnamefont {Nishio-Hamane}}, \bibinfo {author}
  {\bibfnamefont {R.}~\bibnamefont {Arita}}, \bibinfo {author} {\bibfnamefont
  {Y.}~\bibnamefont {Otani}}, \ and\ \bibinfo {author} {\bibfnamefont
  {S.}~\bibnamefont {Nakatsuji}},\ }\bibfield  {title} {\enquote {\bibinfo
  {title} {{Large anomalous Nernst effect at room temperature in a chiral
  antiferromagnet}},}\ }\href {\doibase 10.1038/nphys4181} {\bibfield
  {journal} {\bibinfo  {journal} {Nature Physics}\ }\textbf {\bibinfo {volume}
  {13}},\ \bibinfo {pages} {1085--1090} (\bibinfo {year} {2017})}\BibitemShut
  {NoStop}%
\bibitem [{\citenamefont {Kuroda}\ \emph {et~al.}(2017)\citenamefont {Kuroda},
  \citenamefont {Tomita}, \citenamefont {Suzuki}, \citenamefont {Bareille},
  \citenamefont {Nugroho}, \citenamefont {Goswami}, \citenamefont {Ochi},
  \citenamefont {Ikhlas}, \citenamefont {Nakayama}, \citenamefont {Akebi},
  \citenamefont {Noguchi}, \citenamefont {Ishii}, \citenamefont {Inami},
  \citenamefont {Ono}, \citenamefont {Kumigashira}, \citenamefont {Varykhalov},
  \citenamefont {Muro}, \citenamefont {Koretsune}, \citenamefont {Arita},
  \citenamefont {Shin}, \citenamefont {Kondo},\ and\ \citenamefont
  {Nakatsuji}}]{Kuroda_2017a}%
  \BibitemOpen
  \bibfield  {author} {\bibinfo {author} {\bibfnamefont {K.}~\bibnamefont
  {Kuroda}}, \bibinfo {author} {\bibfnamefont {T.}~\bibnamefont {Tomita}},
  \bibinfo {author} {\bibfnamefont {M.-T.}\ \bibnamefont {Suzuki}}, \bibinfo
  {author} {\bibfnamefont {C.}~\bibnamefont {Bareille}}, \bibinfo {author}
  {\bibfnamefont {A.~A.}\ \bibnamefont {Nugroho}}, \bibinfo {author}
  {\bibfnamefont {P.}~\bibnamefont {Goswami}}, \bibinfo {author} {\bibfnamefont
  {M.}~\bibnamefont {Ochi}}, \bibinfo {author} {\bibfnamefont {M.}~\bibnamefont
  {Ikhlas}}, \bibinfo {author} {\bibfnamefont {M.}~\bibnamefont {Nakayama}},
  \bibinfo {author} {\bibfnamefont {S.}~\bibnamefont {Akebi}}, \bibinfo
  {author} {\bibfnamefont {R.}~\bibnamefont {Noguchi}}, \bibinfo {author}
  {\bibfnamefont {R.}~\bibnamefont {Ishii}}, \bibinfo {author} {\bibfnamefont
  {N.}~\bibnamefont {Inami}}, \bibinfo {author} {\bibfnamefont
  {K.}~\bibnamefont {Ono}}, \bibinfo {author} {\bibfnamefont {H.}~\bibnamefont
  {Kumigashira}}, \bibinfo {author} {\bibfnamefont {A.}~\bibnamefont
  {Varykhalov}}, \bibinfo {author} {\bibfnamefont {T.}~\bibnamefont {Muro}},
  \bibinfo {author} {\bibfnamefont {T.}~\bibnamefont {Koretsune}}, \bibinfo
  {author} {\bibfnamefont {R.}~\bibnamefont {Arita}}, \bibinfo {author}
  {\bibfnamefont {S.}~\bibnamefont {Shin}}, \bibinfo {author} {\bibfnamefont
  {T.}~\bibnamefont {Kondo}}, \ and\ \bibinfo {author} {\bibfnamefont
  {S.}~\bibnamefont {Nakatsuji}},\ }\bibfield  {title} {\enquote {\bibinfo
  {title} {{Evidence for magnetic Weyl fermions in a correlated metal}},}\
  }\href {\doibase 10.1038/nmat4987} {\bibfield  {journal} {\bibinfo  {journal}
  {Nature Materials}\ }\textbf {\bibinfo {volume} {16}},\ \bibinfo {pages}
  {1090--1095} (\bibinfo {year} {2017})}\BibitemShut {NoStop}%
\bibitem [{\citenamefont {Higo}\ \emph {et~al.}(2018)\citenamefont {Higo},
  \citenamefont {Man}, \citenamefont {Gopman}, \citenamefont {Wu},
  \citenamefont {Koretsune}, \citenamefont {van~'t Erve}, \citenamefont
  {Kabanov}, \citenamefont {Rees}, \citenamefont {Li}, \citenamefont {Suzuki},
  \citenamefont {Patankar}, \citenamefont {Ikhlas}, \citenamefont {Chien},
  \citenamefont {Arita}, \citenamefont {Shull}, \citenamefont {Orenstein},\
  and\ \citenamefont {Nakatsuji}}]{Higo_2018a}%
  \BibitemOpen
  \bibfield  {author} {\bibinfo {author} {\bibfnamefont {T.}~\bibnamefont
  {Higo}}, \bibinfo {author} {\bibfnamefont {H.}~\bibnamefont {Man}}, \bibinfo
  {author} {\bibfnamefont {D.~B.}\ \bibnamefont {Gopman}}, \bibinfo {author}
  {\bibfnamefont {L.}~\bibnamefont {Wu}}, \bibinfo {author} {\bibfnamefont
  {T.}~\bibnamefont {Koretsune}}, \bibinfo {author} {\bibfnamefont {O.~M.~J.}\
  \bibnamefont {van~'t Erve}}, \bibinfo {author} {\bibfnamefont {Y.~P.}\
  \bibnamefont {Kabanov}}, \bibinfo {author} {\bibfnamefont {D.}~\bibnamefont
  {Rees}}, \bibinfo {author} {\bibfnamefont {Y.}~\bibnamefont {Li}}, \bibinfo
  {author} {\bibfnamefont {M.-T.}\ \bibnamefont {Suzuki}}, \bibinfo {author}
  {\bibfnamefont {S.}~\bibnamefont {Patankar}}, \bibinfo {author}
  {\bibfnamefont {M.}~\bibnamefont {Ikhlas}}, \bibinfo {author} {\bibfnamefont
  {C.~L.}\ \bibnamefont {Chien}}, \bibinfo {author} {\bibfnamefont
  {R.}~\bibnamefont {Arita}}, \bibinfo {author} {\bibfnamefont {R.~D.}\
  \bibnamefont {Shull}}, \bibinfo {author} {\bibfnamefont {J.}~\bibnamefont
  {Orenstein}}, \ and\ \bibinfo {author} {\bibfnamefont {S.}~\bibnamefont
  {Nakatsuji}},\ }\bibfield  {title} {\enquote {\bibinfo {title} {{Large
  magneto-optical Kerr effect and imaging of magnetic octupole domains in an
  antiferromagnetic metal}},}\ }\href {\doibase 10.1038/s41566-017-0086-z}
  {\bibfield  {journal} {\bibinfo  {journal} {Nature Photonics}\ }\textbf
  {\bibinfo {volume} {12}},\ \bibinfo {pages} {73--78} (\bibinfo {year}
  {2018})}\BibitemShut {NoStop}%
\bibitem [{\citenamefont {Tsai}\ \emph {et~al.}(2020)\citenamefont {Tsai},
  \citenamefont {Higo}, \citenamefont {Kondo}, \citenamefont {Nomoto},
  \citenamefont {Sakai}, \citenamefont {Kobayashi}, \citenamefont {Nakano},
  \citenamefont {Yakushiji}, \citenamefont {Arita}, \citenamefont {Miwa},
  \citenamefont {Otani},\ and\ \citenamefont {Nakatsuji}}]{Tsai_2020}%
  \BibitemOpen
  \bibfield  {author} {\bibinfo {author} {\bibfnamefont {H.}~\bibnamefont
  {Tsai}}, \bibinfo {author} {\bibfnamefont {T.}~\bibnamefont {Higo}}, \bibinfo
  {author} {\bibfnamefont {K.}~\bibnamefont {Kondo}}, \bibinfo {author}
  {\bibfnamefont {T.}~\bibnamefont {Nomoto}}, \bibinfo {author} {\bibfnamefont
  {A.}~\bibnamefont {Sakai}}, \bibinfo {author} {\bibfnamefont
  {A.}~\bibnamefont {Kobayashi}}, \bibinfo {author} {\bibfnamefont
  {T.}~\bibnamefont {Nakano}}, \bibinfo {author} {\bibfnamefont
  {K.}~\bibnamefont {Yakushiji}}, \bibinfo {author} {\bibfnamefont
  {R.}~\bibnamefont {Arita}}, \bibinfo {author} {\bibfnamefont
  {S.}~\bibnamefont {Miwa}}, \bibinfo {author} {\bibfnamefont {Y.}~\bibnamefont
  {Otani}}, \ and\ \bibinfo {author} {\bibfnamefont {S.}~\bibnamefont
  {Nakatsuji}},\ }\bibfield  {title} {\enquote {\bibinfo {title} {Electrical
  manipulation of a topological antiferromagnetic state},}\ }\href {\doibase
  10.1038/s41586-020-2211-2} {\bibfield  {journal} {\bibinfo  {journal}
  {Nature}\ }\textbf {\bibinfo {volume} {580}},\ \bibinfo {pages} {1--6}
  (\bibinfo {year} {2020})}\BibitemShut {NoStop}%
\bibitem [{\citenamefont {Kr{\'{e}}n}\ \emph {et~al.}(1975)\citenamefont
  {Kr{\'{e}}n}, \citenamefont {Paitz}, \citenamefont {Zimmer},\ and\
  \citenamefont {Zsoldos}}]{Kren_1975}%
  \BibitemOpen
  \bibfield  {author} {\bibinfo {author} {\bibfnamefont {E.}~\bibnamefont
  {Kr{\'{e}}n}}, \bibinfo {author} {\bibfnamefont {J.}~\bibnamefont {Paitz}},
  \bibinfo {author} {\bibfnamefont {G.}~\bibnamefont {Zimmer}}, \ and\ \bibinfo
  {author} {\bibfnamefont {{\'{E}}.}~\bibnamefont {Zsoldos}},\ }\bibfield
  {title} {\enquote {\bibinfo {title} {Study of the magnetic phase
  transformation in the mn$_3$sn phase},}\ }\href {\doibase
  10.1016/0378-4363(75)90066-2} {\bibfield  {journal} {\bibinfo  {journal}
  {Physica B+C}\ }\textbf {\bibinfo {volume} {80}},\ \bibinfo {pages}
  {226--230} (\bibinfo {year} {1975})}\BibitemShut {NoStop}%
\bibitem [{\citenamefont {Cable}, \citenamefont {Wakabayashi},\ and\
  \citenamefont {Radhakrishna}(1993)}]{Cable_1993a}%
  \BibitemOpen
  \bibfield  {author} {\bibinfo {author} {\bibfnamefont {J.}~\bibnamefont
  {Cable}}, \bibinfo {author} {\bibfnamefont {N.}~\bibnamefont {Wakabayashi}},
  \ and\ \bibinfo {author} {\bibfnamefont {P.}~\bibnamefont {Radhakrishna}},\
  }\bibfield  {title} {\enquote {\bibinfo {title} {{A neutron study of the
  magnetic structure of Mn$_3$Sn}},}\ }\href {\doibase
  10.1016/0038-1098(93)90400-h} {\bibfield  {journal} {\bibinfo  {journal}
  {Solid State Communications}\ }\textbf {\bibinfo {volume} {88}},\ \bibinfo
  {pages} {161--166} (\bibinfo {year} {1993})}\BibitemShut {NoStop}%
\bibitem [{\citenamefont {Sung}\ \emph {et~al.}(2018)\citenamefont {Sung},
  \citenamefont {Ronning}, \citenamefont {Thompson},\ and\ \citenamefont
  {Bauer}}]{Sung_2018}%
  \BibitemOpen
  \bibfield  {author} {\bibinfo {author} {\bibfnamefont {N.~H.}\ \bibnamefont
  {Sung}}, \bibinfo {author} {\bibfnamefont {F.}~\bibnamefont {Ronning}},
  \bibinfo {author} {\bibfnamefont {J.~D.}\ \bibnamefont {Thompson}}, \ and\
  \bibinfo {author} {\bibfnamefont {E.~D.}\ \bibnamefont {Bauer}},\ }\bibfield
  {title} {\enquote {\bibinfo {title} {{Magnetic phase dependence of the
  anomalous Hall effect in Mn$_3$Sn single crystals}},}\ }\href {\doibase
  10.1063/1.5021133} {\bibfield  {journal} {\bibinfo  {journal} {Applied
  Physics Letters}\ }\textbf {\bibinfo {volume} {112}} (\bibinfo {year}
  {2018}),\ 10.1063/1.5021133}\BibitemShut {NoStop}%
\bibitem [{\citenamefont {Song}\ \emph {et~al.}(2020)\citenamefont {Song},
  \citenamefont {Hao}, \citenamefont {Wang}, \citenamefont {Zhang},
  \citenamefont {Huang}, \citenamefont {Xing},\ and\ \citenamefont
  {Chen}}]{Song_2020}%
  \BibitemOpen
  \bibfield  {author} {\bibinfo {author} {\bibfnamefont {Y.}~\bibnamefont
  {Song}}, \bibinfo {author} {\bibfnamefont {Y.}~\bibnamefont {Hao}}, \bibinfo
  {author} {\bibfnamefont {S.}~\bibnamefont {Wang}}, \bibinfo {author}
  {\bibfnamefont {J.}~\bibnamefont {Zhang}}, \bibinfo {author} {\bibfnamefont
  {Q.}~\bibnamefont {Huang}}, \bibinfo {author} {\bibfnamefont
  {X.}~\bibnamefont {Xing}}, \ and\ \bibinfo {author} {\bibfnamefont
  {J.}~\bibnamefont {Chen}},\ }\bibfield  {title} {\enquote {\bibinfo {title}
  {{Complicated magnetic structure and its strong correlation with the
  anomalous Hall effect in Mn$_3$Sn}},}\ }\href {\doibase
  10.1103/PhysRevB.101.144422} {\bibfield  {journal} {\bibinfo  {journal}
  {Physical Review B}\ }\textbf {\bibinfo {volume} {144422}},\ \bibinfo {pages}
  {1--7} (\bibinfo {year} {2020})}\BibitemShut {NoStop}%
\bibitem [{Note1()}]{Note1}%
  \BibitemOpen
  \bibinfo {note} {M. Ikhlas and S. Nakatsuji, in preparation.}\BibitemShut
  {Stop}%
\bibitem [{\citenamefont {Hicks}\ \emph {et~al.}(2014)\citenamefont {Hicks},
  \citenamefont {Barber}, \citenamefont {Edkins}, \citenamefont {Brodsky},\
  and\ \citenamefont {Mackenzie}}]{Hicks_2014a}%
  \BibitemOpen
  \bibfield  {author} {\bibinfo {author} {\bibfnamefont {C.~W.}\ \bibnamefont
  {Hicks}}, \bibinfo {author} {\bibfnamefont {M.~E.}\ \bibnamefont {Barber}},
  \bibinfo {author} {\bibfnamefont {S.~D.}\ \bibnamefont {Edkins}}, \bibinfo
  {author} {\bibfnamefont {D.~O.}\ \bibnamefont {Brodsky}}, \ and\ \bibinfo
  {author} {\bibfnamefont {A.~P.}\ \bibnamefont {Mackenzie}},\ }\bibfield
  {title} {\enquote {\bibinfo {title} {{Piezoelectric-based apparatus for
  strain tuning}},}\ }\href {\doibase 10.1063/1.4881611} {\bibfield  {journal}
  {\bibinfo  {journal} {Review of Scientific Instruments}\ }\textbf {\bibinfo
  {volume} {85}},\ \bibinfo {pages} {065003} (\bibinfo {year}
  {2014})}\BibitemShut {NoStop}%
\bibitem [{\citenamefont {Duan}\ \emph {et~al.}(2015)\citenamefont {Duan},
  \citenamefont {Ren}, \citenamefont {Liu}, \citenamefont {Li}, \citenamefont
  {Liu},\ and\ \citenamefont {Zhang}}]{Duan_2015a}%
  \BibitemOpen
  \bibfield  {author} {\bibinfo {author} {\bibfnamefont {T.~F.}\ \bibnamefont
  {Duan}}, \bibinfo {author} {\bibfnamefont {W.~J.}\ \bibnamefont {Ren}},
  \bibinfo {author} {\bibfnamefont {W.~L.}\ \bibnamefont {Liu}}, \bibinfo
  {author} {\bibfnamefont {S.~J.}\ \bibnamefont {Li}}, \bibinfo {author}
  {\bibfnamefont {W.}~\bibnamefont {Liu}}, \ and\ \bibinfo {author}
  {\bibfnamefont {Z.~D.}\ \bibnamefont {Zhang}},\ }\bibfield  {title} {\enquote
  {\bibinfo {title} {{Magnetic anisotropy of single-crystalline Mn$_3$Sn in
  triangular and helix-phase states}},}\ }\href {\doibase 10.1063/1.4929447}
  {\bibfield  {journal} {\bibinfo  {journal} {Applied Physics Letters}\
  }\textbf {\bibinfo {volume} {107}},\ \bibinfo {pages} {082403} (\bibinfo
  {year} {2015})}\BibitemShut {NoStop}%
\bibitem [{\citenamefont {Staunton}\ \emph {et~al.}(2014)\citenamefont
  {Staunton}, \citenamefont {Banarjee}, \citenamefont {Dias}, \citenamefont
  {Deak},\ and\ \citenamefont {Szunyogh}}]{Staunton_2014}%
  \BibitemOpen
  \bibfield  {author} {\bibinfo {author} {\bibfnamefont {J.~B.}\ \bibnamefont
  {Staunton}}, \bibinfo {author} {\bibfnamefont {R.}~\bibnamefont {Banarjee}},
  \bibinfo {author} {\bibfnamefont {M.~D.~S.}\ \bibnamefont {Dias}}, \bibinfo
  {author} {\bibfnamefont {A.}~\bibnamefont {Deak}}, \ and\ \bibinfo {author}
  {\bibfnamefont {L.}~\bibnamefont {Szunyogh}},\ }\bibfield  {title} {\enquote
  {\bibinfo {title} {{Fluctuating local moments, itinerant electrons, and the
  magnetocaloric effect: Compositional hypersensitivity of FeRh}},}\ }\href
  {\doibase 10.1103/PhysRevB.89.054427} {\bibfield  {journal} {\bibinfo
  {journal} {Physical Review B}\ }\textbf {\bibinfo {volume} {89}},\ \bibinfo
  {pages} {1--7} (\bibinfo {year} {2014})}\BibitemShut {NoStop}%
\end{thebibliography}%

\end{document}